\documentclass[aps,superscriptaddress,twocolumn,amsmath,amssymb]{revtex4-1}

\usepackage{dsfont,amsthm,amsbsy}
\usepackage{amssymb}
\usepackage{amsmath}
\usepackage{bbm}
\usepackage{graphicx}
\usepackage{epstopdf}
\usepackage{subfigure}
\usepackage{natbib}
\usepackage{epsfig}
\usepackage{amsfonts}
\usepackage{mathrsfs}
\usepackage{sidecap}
\usepackage{lipsum}
\usepackage{xcolor}
\usepackage[toc,page,title,titletoc,header]{appendix}
\usepackage[colorlinks,linkcolor=blue,citecolor=blue,anchorcolor=blue, urlcolor=blue]{hyperref}
\usepackage{hyperref}
\usepackage{resizegather}
\usepackage{float}
\usepackage{mathbbol}
\usepackage[normalem]{ulem}
\usepackage{cancel}
\usepackage{upgreek}
\usepackage{graphics,dcolumn,bm}
\usepackage{rotate,color}
\usepackage{times}
\newcommand\redsout{\bgroup\markoverwith{\textcolor{red}{\rule[0.5ex]{2pt}{0.4pt}}}\ULon}

\newcommand{\bl}{\begin{aligned}}
\newcommand{\el}{\end{aligned}}
\def\be{\begin{equation}}
\def\ee{\end{equation}}

\def\bi{\begin{itemize}}
\def\ei{\end{itemize}}
\def\bn{\begin{enumerate}}
\def\en{\end{enumerate}}
\def\bea{\begin{eqnarray}}
\def\eea{\end{eqnarray}}
\def\no{\nonumber}
\def\ba{\begin{array}}
\def\ea{\end{array}}
\def\bd{\begin{displaymath}}
\def\ed{\end{displaymath}}

\begin{document}

\title{Floquet dynamical quantum phase transitions under synchronized periodic driving}

\author{R. Jafari}
\email[]{jafari@iasbs.ac.ir, rohollah.jafari@gmail.com (corresponding author)}
\affiliation{Department of Physics, Institute for Advanced Studies in Basic Sciences (IASBS), Zanjan 45137-66731, Iran}
\affiliation{Department of Physics, University of Gothenburg, SE 412 96 Gothenburg, Sweden}
\affiliation{Beijing Computational Science Research Center, Beijing 100094, China}

\author{Alireza Akbari}
%\email[]{}
\affiliation{Max Planck Institute for the Chemical Physics of Solids, D-01187 Dresden, Germany}
\affiliation{Max Planck POSTECH Center for Complex Phase Materials and Department of Physics, POSTECH, Pohang, Gyeongbuk 790-784, Korea}

\author{Utkarsh Mishra}
%\email{}
\affiliation{University of Electronic Science and Technology of China, Chengdu 610051, PR China}

\author{Henrik Johannesson}
%\email[]{}
\affiliation{Department of Physics, University of Gothenburg, SE 412 96 Gothenburg, Sweden}
%\date{\today}

\begin{abstract}
We study a generic class of fermionic two-band models under synchronized periodic driving, i.e., with the different terms in a Hamiltonian subject to periodic drives with the same {\color{black} frequency and phase}. With all modes initially in a maximally mixed state, the synchronized drive is found to produce nonperiodic patterns of dynamical quantum phase transitions, with their appearance determined by an interplay of the band structure and the frequency of the drive. A case study of the anisotropic XY chain in a transverse magnetic field, transcribed to an effective two-band model, shows that the modes come with quantized geometric phases, allowing for the construction of an effective dynamical order parameter. Numerical studies in the limit of a strong magnetic field reveal distinct signals of precursors of dynamical quantum phase transitions also when the initial state of the XY chain is perturbed slightly away from maximal mixing, suggesting that the transitions may be accessible experimentally. A blueprint for an experiment built around laser-trapped circular Rydberg atoms is proposed.
\end{abstract}


\maketitle

\section{Introduction\label{SE1}}

Dynamical phase transitions $-$ phase transitions away from equilibrium $-$ span a large spectrum of phenomena, from the domain formation in the early universe~\cite{Kibble1976} to abrupt changes in the relaxation dynamics of strongly correlated quantum many-particle systems~\cite{Eckstein2009}.  A particularly intriguing variety are those nonequilibrium phase transitions where physical quantities become nonanalytic as a function of time, known as {\em dynamical quantum phase transitions} (DQPTs)~\cite{Heyl2013}.

The notion of a DQPT
is rooted in the formal similarity between an equilibrium partition function and the Loschmidt amplitude that measures the overlap between an initial and time-evolved state of a system out of equilibrium~\cite{Heyl2018,Zvyagin2016}. When two such states become orthogonal, the vanishing of the Loschmidt amplitude causes a nonanalyticity in the associated dynamical free energy, mimicking the behavior of the free energy (ground state energy) of an equilibrium system at a thermal (quantum) phase transition. While being conceptually rather different phenomena, this formal analogy has informed the study of DQPTs, allowing concepts and ideas to be borrowed from the theory of equilibrium phase transitions. By now, there is a large literature on the theory~\cite{Heyl2013,Karrasch2013,Kriel2014,Canovi2014,Andraschko2014,vajna2014disentangling,Heyl2014,Heyl2015,vajna2015topological,Budich2016,Huang2016,
Bhattacharya2017,Heyl2017,bhattacharya2017emergent,zhou2018dynamical,Lang2018,Zhou2019,Mendl2019,Uhrich2020,Nicola2021,Bandyopadhyay2021,Jafari2021,Halimeh2021, Sacramento2021} and modeling of DQPTs, with a growing number of experimental reports~\cite{Jurcevic2017,Zhang2017,Bernien2017,Flaschner2018,YangTianYang2019,Guo2019,wang2019simulating,Tian2019,Tian2020,Nie2020,Chen2020}.

Most works so far have focused on DQPTs triggered by a quantum quench where a system is forced out of equilibrium by a sudden change of its Hamiltonian. More recently, however, there have been studies of DQPTs also in periodically driven systems~\cite{Oka2019} $-$ known as {\em Floquet DQPTs} $-$ including those which exhibit nonequilibrium phases like Floquet topological phases~\cite{Lindner2011} and discrete time crystals~\cite{Else2016}. These systems have been found to display DQPTs with non-decaying return probabilities which should make them easier to monitor in the laboratory. In contrast, DQPTs set off by a single quench have decaying return probabilities and are therefore observable only on transient time scales.

The periodic drives that have been studied up till now, supporting Floquet DQPTs, have come in various types: schemes that can be described by a time-independent Hamiltonian in a rotating frame~\cite{yang2019floquet,Zamani2020,Jafari2020a}; stroboscopic drives, effectively emulating a periodically repeated quantum quench~\cite{Zhou2021,Hamazaki2021}; or a single quench in a periodically driven discrete time crystal~\cite{kosior2018dynamical,kosior2018dynamicalb}. Considering the potential for future experimental studies of DQPTs, it is expedient to inquire about other types of periodic driving schemes. Specifically, do other schemes support a classification of the nonequilibrium phases in terms of a topological order parameter, known from most quench and periodically driven systems studied so far? How does the choice of initial states and tuning of parameters impact the Floquet DQPTs? And what about the generalization of Floquet DQPTs from pure states to density matrices, allowing a system to be entangled with its environment? These are some of the questions we wish to address in this paper.

For this purpose we introduce a class of periodically driven lattice models in one dimension (1D) which {\em cannot} be mimicked by a stroboscopically driven Hamiltonian or by a Hamiltonian which becomes time-independent in a rotating frame.  A minimal setup, amenable to exact analysis, is a noninteracting fermionic two-band model with {\em all} terms in its Hamiltonian subject to periodic drives with the same {\color{black} frequency and phase} $-$ hence the notion of {\em synchronized periodic driving}. As we will show, hiding behind the apparent triviality of such a model is some very interesting physics, showcasing a nonperiodic sequence of DQPTs {\color{black} when properly initialized}, and with the DQPTs caused by single modes with quantized geometric phases. Also, different from expectations~\cite{yang2019floquet,Zhou2021}, the associated time-independent Floquet Hamiltonian that governs the stroboscopic time evolution over one cycle of the drive does not necessarily support topologically nontrivial phases: DQPTs may appear independent of the topological features of the ground state of the Floquet Hamiltonian, similar to the appearance of ``accidental" DQPTs of time-independent models subject to a single quantum quench~\cite{vajna2015topological,Budich2016,Huang2016}.

The paper is structured as follows: In the section to follow we introduce the notion of a synchronized periodic drive in the setting of noninteracting fermionic two-band models. We briefly review some pertinent key concepts in the theory of DQPTs and show how such transitions are triggered in these models given a synchronized drive with all modes initially populating the two bands with equal probability (``maximal mixing" {\color{black} in the language of an open system}). In Sec. III we analyze a benchmark model $-$ the anisotropic XY chain in a transverse magnetic field $-$ by applying the formalism from Sec. II after having fermionized the model, in this way reducing it to a noninteracting two-band Hamiltonian. {\color{black} Numerical data for the rate function of the Loschmidt echo and an effective dynamical order parameter exhibit clear signals of precursors of DQPTs also when the initial state of the system is perturbed slightly off maximal mixing, suggesting that DQPTs under synchronized periodic driving may be accessible experimentally.}  In Sec. IV we {\color{black} follow up on this and} sketch the backbones of an experimental protocol to realize DQPTs in the synchronously driven XY chain studied in the preceding section. Sec. V, finally, contains a brief summary and outlook. Throughout the paper we use units with $\hbar = k_B = 1$.

\section{Fermionic two-band models under synchronized periodic driving\label{SE2}}

As a preamble, consider the time-independent first-quantized Hamiltonian $H\!=\! H(\{J_{\ell}\}_{\ell =1}^m)$, where $J_{\ell}$ is the amplitude of the $\ell$:th term in $H$. Subjecting {\em all} terms in $H$ to time-dependent drives $\lambda_\ell(t) (\ell=1,2,\ldots,m$) with the same frequency and phase, we call the resulting time-dependent Hamiltonian {\em synchronously driven} (not to be confused with the phenomenon of "synchronization" where an oscillating system with a stable limit cycle phase locks to an external signal or to another oscillating system to which it is coupled). For simplicity we shall assume that the amplitudes of the drives for the different terms are also the same, i.e., $\lambda_1(t) = \lambda_2(t) = \cdots = \lambda(t)$, with $\lambda(t)$ a real dimensionless scalar function which we take to satisfy {\color{black} $\lambda(0) > 0$}. In this case the resulting time-dependent Hamiltonian $H(t)$ is simply given by $H(t) = \lambda(t)H$.

Initializing the system in one of the eigenstates $|\varphi_n\rangle$ of $H$, the time-evolved state $|\varphi_{n}(t)\rangle$ governed by $H(t)$ is readily obtained from the time-dependent Schr\"odinger equation, yielding
%
%%%%%%%%%%%%%%%%%%%%%%%%%%%%%%%%%%%%%%%  Eq. 1  %%%%%%%%%%%%%%%%%%%%%%%%%%%%%%%%%%%%%%%%%%%%%
\begin{equation} \label{eq:1}
|\varphi_{n}(t)\rangle=e^{-i\varepsilon_{n}\tau(t)}|\varphi_{n}\rangle,
\end{equation}
%%%%%%%%%%%%%%%%%%%%%%%%%%%%%%%%%%%%%%%%%%%%%%%%%%%%%%%%%%%%%%%%%%%%%%%%%%%%%%%%%%%%
%
where $\varepsilon_n$ is the eigenvalue of $H$ associated with the state $|\varphi_n\rangle$, and with $\tau(t)=\int_{0}^{t}\lambda(t')dt'$. As expected, the synchronously driven system hence remains in the same physical state under time evolution. One may thus be tempted to dismiss a setup with synchronized driving as being trivial. While this is certainly so
when the initial state is an eigenstate of $H$, the situation changes when this condition is removed.

To see the implications, we shall consider the mode decomposition of a generic fermionic two-band model with Hamiltonian ${\cal H}$, subject to a {\em periodic} synchronized drive $\lambda(t)$, and with one mode initialized in a superposition of two eigenstates.  Examples of ${\cal H}$ include band insulators and
mean-field superconductors, paradigmatic in the study of symmetry-protected topological phases~\cite{Chiu2016}.
Other noted instances are effective models of spin chains, including the {\color{black} transverse field} Ising and XY chains~\cite{Franchini2017}, {\color{black} to be discussed in the next section}. Such two-band systems are widely studied in the literature, and several of them have been realized experimentally, in condensed matter or with analog quantum simulators using ultracold atoms in optical lattices~\cite{Lewenstein2012}.

Assuming periodic boundary conditions, we write the Fourier transformed Hamiltonian ${\cal H}$ in second quantization as
%%%%%%%%%%%%%%%%%%%%
\begin{equation} \label{modedecomp}
{\cal H} =\sum_{k} {\boldsymbol{c}}^{\dagger}_{k}H_{k}{\boldsymbol{c}}_{k},
\end{equation}
%%%%%%%%%%%%%%%%%%%%
where, for simplicity, and with an eye to applications to come, we assume ${\cal H}$ to be defined in 1D.
Depending on the particular model, ${\boldsymbol{c}}^{\dagger}_{k}$ is an ordinary two-spinor, ${\boldsymbol{c}}^{\dagger}_{k}\!=\!(c_{k,A}^{\dagger},c_{k,B}^{\dagger})$, with $c_{k,A/B}^{\dagger}$ fermion creation operators where $A$, $B$ refer to two internal degrees of freedom (like spin or sublattice), or a Nambu spinor, ${\boldsymbol{c}}^{\dagger}_{k}\!=\!(c_{k}^{\dagger},~c_{-k})$, appropriate for superconductors and certain fermionized models of spin chains. In the first (second) case, the sum in Eq. (\ref{modedecomp}) runs over all momenta in the Brillouin zone, $k \in [-\pi, \pi]$ (half the Brillouin zone, $k \in [0, \pi]$). $H_{k}$ is the first-quantized Hamiltonian for a mode with momentum $k$, with $H_k(t)=\lambda(t)H_{k}$ the corresponding synchronously driven mode Hamiltonian. {\color{black} We here focus on the generic class of two-band models where}
%%%%%%%%%%%
\bea
\label{eq:Hk0}
H_{k}= {\color{black} \frac{\Delta_k}{2}\boldsymbol{n}(k) \cdot \boldsymbol{\sigma},}
\eea
%%%%%%%%%%%
{\color{black} with $\Delta_k$ the gap between the two eigenstates of $H_k$, $\boldsymbol{n}_k \!=\! \left(\sin \theta_{k}\cos \phi_{k},\sin \theta_{k}\sin \phi_{k},\cos \theta_{k} \right)$ a unit vector with $\theta_k$ and $\phi_k$ spherical coordinates, and with $\boldsymbol{\sigma} = (\sigma_x, \sigma_y, \sigma_z)$ the vector of Pauli matrices. Explicitly,} $\Delta_k = 2\varepsilon_k$, with $\pm \varepsilon_k = \varepsilon_k^{\pm}$ the two eigenvalues of $H_{k}$ corresponding to the eigenstates
%
%%%%%%%%%%%%%%%%%%%%%%%%%%%%%%%%%%%%  Eq. 6   %%%%%%%%%%%%%%%%%%%%%%%%%%%%%%%%%%%%%%%%%%%%%%
\bea
\label{eigenstates}
|\varphi^{+}_{k}\rangle\!=\!\left(\!
                          \begin{array}{c}
                            \mbox{e}^{-{\it i}\phi_{k}}\!\cos(\frac{\theta_{k}}{2}) \\
                            \sin(\frac{\theta_{k}}{2}) \\
                          \end{array}
                        \!\right), \
|\varphi^{-}_{k}\rangle\!=\!\left(\!
                          \begin{array}{c}
                            -\mbox{e}^{-{\it i}\phi_{k}}\!\sin(\frac{\theta_{k}}{2}) \\
                            \cos(\frac{\theta_{k}}{2}). \\
                          \end{array}
                        \!\right)
                        . \quad
\eea
%%%%%%%%%%%%%%%%%%%%%%%%%%%%%%%%%%%%%%%%%%%%%%%%%%%%%%%%%%%%%%%%%%%%%%%%%%%%%%%%%%

The time evolution of $|\varphi^{\pm}_{k}\rangle$ can be read off from Eq. (\ref{eq:1}), yielding $|\varphi^{\pm}_{k}(t)\rangle=e^{\mp i\varepsilon_{k}\tau(t)}|\varphi^{\pm}_{k}\rangle$. It follows that the time evolution of a superposition $|\varphi_{k}\rangle = \alpha_{k}^{-}|\varphi_{k}^{-}\rangle+\alpha_{k}^{+}|\varphi_{k}^{+}\rangle$ takes the form
%%%%%%%%%%%%%%%%%%%%%%%%%%%%%%%%%%%%  Eq. 7   %%%%%%%%%%%%%%%%%%%%%%%%%%%%%%%%%%%%%%%%%%%%%%
\bea
\label{eq:super}
|\varphi_{k}(t)\rangle=\alpha_{k}^{-}e^{i\varepsilon_{k}\tau(t)}|\varphi_{k}^{-}\rangle+\alpha_{k}^{+}e^{- i\varepsilon_{k}\tau(t)}|\varphi_{k}^{+}\rangle,
\eea
%%%%%%%%%%%%%%%%%%%%%%%%%%%%%%%%%%%%%%%%%%%%%%%%%%%%%%%%%%%%%%%%%%%%%%%%%%%%%%%%%%
with $|\alpha_{k}^{-}|^{2}+|\alpha_{k}^{+}|^{2}=1$. While this is an entirely expected result given Eq. (\ref{eq:1}), the expression in (\ref{eq:super}) contains the seed for generating DQPTs of an
unconventional brand, different from those studied so far in periodically driven systems. In the following section we
shall see how this comes about.

\subsection{Dynamical quantum phase transitions for pure states\label{SE2a}}

The key concept in the theory of DQPTs~\cite{Heyl2018,Zvyagin2016} is the Loschmidt amplitude
%%%%%%%%%%%%%%%%%%%%%%%%%%%%%%%%%%
\begin{equation} \label{eq:Lamplitude}
{\cal G}(t) = \langle \psi(0)|\psi(t)\rangle ,
\end{equation}
%%%%%%%%%%%%%%%%%%%%%%%%%%%%%%%%%
which measures the overlap of a state $|\psi(0)\rangle$ at time $t=0$ with its time-evolved descendant $|\psi(t)\rangle$. When the Hamiltonian as here is synchronously driven, we have that
\begin{equation} \label{evolution}
\langle \psi(0)|\psi(t)\rangle = \langle \psi(0)| e^{-i H \tau(t)} | \psi(0) \rangle,
\end{equation}
%%%%%%%%%%%%%%%%%%%%%%%%%%%%%%%%%
with $\tau(t)$ introduced in Eq. (\ref{eq:1}). The return probability ${\cal L}(t) \equiv |{\cal G}(t)|^2$ linked with the amplitude ${\cal G}(t)$ is the Loschmidt echo, sometimes also termed ``dynamical fidelity".

A DQPT is signaled by the vanishing of ${\cal G}(t)$, causing a nonanalyticity in the rate function $g_{\cal G}(t)$ in the thermodynamic limit,
%%%%%%%%%%%%%%%%%%%%%%%%%%%%%%%%
\begin{equation} \label{eq:rate}
g_{\cal G}(t) = - \lim_{N \rightarrow \infty} N^{-1} \ln {\cal G}(t),
\end{equation}
%%%%%%%%%%%%%%%%%%%%%%%%%%%%%%%%
with $N$ the number of degrees of freedom. The rate function $g_{\cal G}(t)$ can be interpreted as a dynamical free energy density, with time $t$ replacing the control parameter at equilibrium (like temperature in a thermally driven phase transition) and with the Loschmidt amplitude ${\cal G}(t)$ masquerading as a partition function (a role supported by the property that a continuation of ${\cal G}(t)$ into the complex plane formally yields a boundary partition function~\cite{Mussardo1995}).

To apply the formalism to our generic two-band model with second-quantized Hamiltonian ${\cal H}$, we first note that the modes in $k$-space are decoupled, implying that all eigenstates $|\psi\rangle$ of ${\cal H}$ can be factorized into those eigenstates of $H_{k}$ which are occupied. Consider, for example, a system initialized in the ground state $|\psi_0\rangle$ of ${\cal H}$, with the lower (upper) band completely filled (empty), i.e., with the system at ``half filling". Reverting to a first-quantized formalism for $|\psi_0\rangle$, we can write
%%%%%%%%%%%%%%%%%%%%%%%%%%%%%%%%
\begin{equation} \label{eq:product}
|\psi_0\rangle = \otimes_k |\varphi_{k}^{-}\rangle.
\end{equation}
%%%%%%%%%%%%%%%%%%%%%%%%%%%%%%%%

Any external drive which induces a time-dependence in $H_{k}$ without coupling the $k$-modes $-$ such as a synchronized periodic driving $-$ will preserve this property, and hence
%%%%%%%%%%%%%%%%%%%%%%%%%%%%%%%
\begin{equation} \label{eq:timeproduct}
|\psi_0(t)\rangle = \otimes_k |\varphi_{k}^{-}(t)\rangle,
\end{equation}
%%%%%%%%%%%%%%%%%%%%%%%%%%%%%%%%
where $|\varphi_{k}^{-}(t)\rangle = e^{i\varepsilon_{k}\tau(t)}|\varphi_{k}^{-}\rangle$ (cf. text after Eq. (\ref{eigenstates})). Thus, the dynamics can be monitored for each mode $k$ separately.

Now suppose that the system is again initialized with all modes in the lower band except for one, say $k^{\prime}$, which is instead put in a superposition $|\varphi_{k^{\prime}}\rangle$ of the lower and upper band as in Eq. (\ref{eq:super}), with
$|\alpha^{-}_{k'}|^2 (|\alpha^{+}_{k'}|^2)$ the probability for occupancy of the lower (upper) band. We call such an initial state $|\psi^{\prime}(0)\rangle$, and write the Loschmidt amplitude on factorized form as
%%%%%%%%%%%%%%%%%%%%%%%%%%%%%%%%%%%%
\begin{eqnarray} \label{eq:G(t)}
{\cal G}(t) &=& \langle \psi^{\prime}(0) | \psi^{\prime}(t) \rangle = \prod_{k\neq k^\prime} \langle \varphi_{k}^{-} | \varphi_{k}^{-}(t) \rangle \times \langle \varphi_{k^\prime} | \varphi_{k^\prime}(t) \rangle \nonumber \\
& = & \prod_{k\neq k^\prime} {\cal G}^{-}_k(t) \times {\cal G}_{k^\prime}(t).
\end{eqnarray}
%%%%%%%%%%%%%%%%%%%%%%%%%%%%%%%%%
As follows from Eq. (\ref{eq:1}), a zero of ${\cal G}(t)$ can only come from a Loschmidt amplitude for a mode which is {\em not} put in an eigenstate of the Hamiltonian, that is, ${\cal G}_{k^\prime}(t) = \langle \varphi_{k^\prime} | \varphi_{k^\prime}(t) \rangle $ in (\ref{eq:G(t)}). {\color{black} As a Gedanken experiment we may envision that we can  
fine tune the amplitudes in Eq. (\ref{eq:super}), choosing $|\alpha^{-}_{k'}|^2 = |\alpha^{+}_{k'}|^2 = 1/2$, i.e., with the mode initially populating the two bands with equal probability. It then follows that
%%%%%%%%%%%%%%%%%%%%%%%%%%%%%
\begin{equation} \label{eq:cosine}
{\cal G}_{k^\prime}(t) = \cos(\varepsilon_{k^{\prime}}\tau(t)),
\end{equation}
%%%%%%%%%%%%%%%%%%%%%%%%%%%%%
with} zeros at instants of time {\color{black} $t^{\ast}_{k^\prime\!,n}$} for which
%%%%%%%%%%%%%%%%%%%%%%%%%%%%%%%%%%%%%%%%%%%%%%
\begin{equation} \label{eq:tau}
{\color{black}  \tau(t^{\ast}_{k^\prime\!,n})} = \frac{\pi}{\varepsilon_{k^{\prime}}}(n+1/2); \;
 \ \ n= 0,1,2,\ldots.
\end{equation}
%%%%%%%%%%%%%%%%%%%%%%%%%%%%%%%%%%%%%%%%%%%%%%
To make explicit that these times are {\em critical}, associated with DQPTs, and also, that they derive from a particular mode $k^\prime$, we have denoted them by $t_{k^{\prime}\!,n}^{\ast}$, with $n=0,1,2,\ldots$. The generalization to the case where more than one mode is put in an initial state with $|\alpha^{-}|^2 \!=\! |\alpha^{+}|^2 \!=\! 1/2$ is immediate: The system will now exhibit DQPTs at several sequences of critical times $t_{k^{\prime}\!,n}^{\ast}$, with $k^{\prime} = k_1, k_2, \ldots$ and with $n= 0,1,2,...$. By an abuse of language we call these modes ``critical" (although their presence does not by itself lead to a DQPT; for this, the thermodynamic limit must also be taken \cite{Heyl2018,Zvyagin2016}). 

The sequences of possible DQPTs thus unveiled have an interesting property.  As an illustration, let us take a drive where $\lambda(t) = \lambda_0 + \lambda_1\cos(\omega t)$ (a case we shall return to), implying that $\tau(t) = \int_0^t \lambda(t^\prime) dt^\prime = \lambda_0 t + (\lambda_1/\omega) \sin(\omega t)$. While the drive is periodic, with period $T=2\pi/\omega$, this is not the case for the critical times $t_{k^{\prime}\!,n}^{\ast}$, which, according to Eq. (\ref{eq:tau}), are {\em nonperiodic} for this choice of drive. This is different from critical times of DQPTs set off by a quantum quench which display a periodicity for each critical mode $k^\prime$, as proved in Ref.~\cite{Huang2016} for any number of bands (where, in the case of several bands, a critical mode is defined by occupying a state with equal amplitudes for all bands). Case studies of periodically driven two-band systems different from the type considered here have also found that DQPT critical times tied to a given critical mode are periodic~\cite{yang2019floquet,Zamani2020,Jafari2020a,Zhou2021,Hamazaki2021,kosior2018dynamical}. Synchronously driven systems provide a counterpoint, demonstrating that a periodicity of such critical times is not an intrinsic feature of DQPTs.

As transpires from our analysis of a generic two-band model, the appearance of DQPTs under synchronized periodic driving is conditioned on the initialization of the system, where one or several $k$-modes are put in a superposition of states in the upper and lower band with equal amplitudes. The presence of such states are necessary for the existence of DQPTs in any two-band system, however, they may appear {\color{black} also} from the dynamics, with no need for fine tuning the initial state. In 1D, when considering quench dynamics, a sufficient condition for the existence of critical modes $k^\prime$ is that the quench is taken between two Hamiltonians with topologically distinct ground states as characterized by their winding numbers~\cite{vajna2015topological}.

{\color{black} Given any} system driven periodically with a frequency $\omega=2\pi/T$, it has been conjectured that critical modes appear dynamically provided that the associated time-independent Floquet Hamiltonian that controls the stroboscopic time evolution (i.e., the time evolution monitored at instants of time $t_n = t + nT, n=0,1,2,\ldots$) supports a topologically nontrivial ground state~\cite{yang2019floquet}. The Floquet Hamiltonian for a synchronously driven systems is trivial to obtain, being equal to the leading term in a Magnus expansion of the full time-dependent Hamiltonian (since no time ordering of the time-evolution operator is necessary, thus making all higher-order terms in the expansion vanish; for details, see Ref.~\cite{Bukov2015}). It follows that the Floquet Hamiltonian $H_{Fk}$ corresponding to the synchronously driven mode Hamiltonian $H_k(t) = \lambda(t)H_{k}$ is trivially given by an average over one period $T$ of the drive:
%%%%%%%%%%%%%%%%%%%%%%%%%%%%%%%%%%%%%%%%%%%
\begin{equation} \label{Floquet}
H_{Fk} = \left( \frac{1}{T} \int_{t}^{t + T} \lambda(t^{\prime}) dt^{\prime} \right) \times H_{k} = \mbox{const.} \times H_{k}.
\end{equation}
%%%%%%%%%%%%%%%%%%%%%%%%%%%%%%%%%%%%%%%%%%%
Now suppose that $H_{k}$, and by that $H_{Fk}$, does support a topologically nontrivial ground state. Does this imply that a critical mode $k^\prime$ is always ensured to appear dynamically, by this making the fine tuning of the initial state unnecessary? The answer is no. For example, suppose that one takes $|\psi_0\rangle$ in Eq. (\ref{eq:product}) as initial state, and that $|\psi_0\rangle$ is topologically nontrivial. Since all $k$-modes occupy eigenstates of $H_{k}$, the synchronously driven time evolution will return these same eigenstates multiplied by a phase, with no critical mode appearing; cf. Eq. (\ref{eq:1}). This shows that a Floquet Hamiltonian with a topologically nontrivial ground state does not necessarily imply the appearance of DQPTs under synchronized periodic driving.

From this one may be led to infer that DQPTs in these systems only show up if the initial state is fine tuned. However, such fine tuning of pure states is experimentally hard to achieve, if at all possible with present-day quantum technology. A possible way out is suggested by the very feature that a critical mode has equal probability to be found in one of two eigenstates. This is as if the mode were occupying a maximally mixed state of a qubit, and such states may be easier to prepare in the laboratory.  We shall return to this issue when discussing experimental realizations in Sec.~IV. For now, and as a forerunner to that discussion, let us briefly review how the pure-state formalism used in this subsection can be generalized to mixed states.
%Recent proposals to create fast scrambling, i.e., a process where local information spreads to other degrees of freedom in a quantum system on experimentally accessible time scales, here holds great %promise. In short, as one approaches the scrambling time, $t_s \sim \ln N$, with $N$ counting the number of coupled qubits in the system and subject to unitary time evolution, the reduced density matrix %of each qubit is expected to approach maximal mixing. We shall return to this type of protocol when discussing experimental realizations in Sec.~V. For now, and as a forerunner to that discussion, let us &briefly review how the pure-state formalism used in this subsection generalizes to mixed states.

\subsection{Dynamical quantum phase transitions for mixed states\label{SE2b}}

Given that the initial state in which a system is prepared in an experiment is typically a mixed state, it is vital to inquire how to handle this more realistic situation. This issue was addressed in general terms by the authors of Refs.~\cite{Bhattacharya2017,Heyl2017}, {\color{black} and we here follow their approach.

One starts by considering the density matrix $\rho(0)$ of the mixed state at $t=0$. Expanding $\rho(0)$ in its eigenbasis $\{|\psi_i\rangle\}$,
\begin{equation} \label{eq:density}
\rho(0) = \sum_i p_i|\psi_i\rangle\langle\psi_i|,
\end{equation}
the purification $|w(0)\rangle$ of the mixed state can be written as
%%%%%%%%%%%%%%%%%%%
\begin{equation} \label{eq:pure}
|w(0)\rangle = \sum_i \sqrt{p_i} |\psi_i\rangle \otimes |\psi^{\prime}_i\rangle_{\text{aux}},
\end{equation}
%%%%%%%%%%%%%%%%%%%
where $\{|\psi^{\prime}_i\rangle_{\text{aux}}\}$ are auxiliary orthonormal states with the property that $\rho(0) = \text{Tr}_{\text{aux}}\Big(|w(0)\rangle\langle w(0)|\Big)$. Introducing the time-evolution operator, given by $U(t,0) = \exp(-iH\tau(t))$ under a synchronized periodic drive, the purification evolves as $|w(t)\rangle =  U(t,0) \otimes {1}_{\text{aux}} |w(0)\rangle$,
with ${1}_{\text{aux}}$ the identity operator in the Hilbert space of the auxiliary states. It follows that the generalized (mixed state) Loschmidt amplitude ${\cal G}^{\rho}(t) = \langle w(0)| w(t)\rangle$ can be written as
%%%%%%%%%%%%%%%%%%%%%%%%%
\begin{equation} \label{eq:genLoschmidt}
{\cal G}^{\rho}(t) = \text{Tr}\Big(\rho(0)U(t,0)\Big).
\end{equation}
%%%%%%%%%%%%%%%%%%%%%%%

Carrying out a mode decomposition, with the mode Hamiltonians $H_k$ expressed in the basis of Eq. (\ref{eq:Hk0}), and assuming that the initial state at $t=0$ is in thermal equilibrium at a temperature $1/\beta$, we have that ${\cal G}^{\rho}(t) = \prod_k {\cal G}^{\rho}_k(t)$ with ${\cal G}^{\rho}_k(t) = \mbox{Tr}\Big(\rho_k(0)U_k(0,t)\Big)$, where
%%%%%%%%%%%%%%%%%%%
\begin{eqnarray} \label{eq:rhok}
\rho_{k}(0) &=& \frac{e^{-\beta H_{k}(0)}}{{\rm Tr}(e^{-\beta H_{k}(0)})} \nonumber \\
&=& \frac{1}{2}\Big(\mathbb{1}-\tanh(\lambda(0)\beta \varepsilon_{k}) \boldsymbol{n}_k \cdot \boldsymbol{\sigma}\Big),
\end{eqnarray}
%%%%%%%%%%%%%%%%%%
with $\lambda(0)$ the initial value of the driving function.

An explicit expression for $U_k(t,0)$ is obtained by solving the equation $i \partial_t{U}_{k}(t)=U_{k}(t,0) H_{k}(t)$, implied by the definition $U_k(t,0) \equiv \exp(-iH_k\tau(t))$. In the next section we shall study an effective fermionic two-band model for a spin chain for which there is no $x$-component in the unit vector $\boldsymbol{n}_k$ that appears in Eqs. (\ref{eq:Hk0}) and (\ref{eq:rhok}). For this class of models one finds that}
%%%%%%%%%%%%%%%%%%
\bea
\label{eq:Uk}
{\color{black} U_{k}(t,0)=\left(
       \begin{array}{cc}
         u_{k,11}(t) &  u_{k,12}(t) \\
         u_{k,21}(t) &  u_{k,22}(t) \\
       \end{array}
     \right)}
\eea
%%%%%%%%%%%%%%%%%
{\color{black} with the simple expressions}
%%%%%%%%%%%%%%%%%%%%%%%%%%%%%%%% Eq. 13 %%%%%%%%%%%%%%%%%%%%%%%%%%%%%%%%
\begin{eqnarray}
\label{eq:Uii}
\bl
u_{k,11}(t)
&=\cos^{2}(\frac{\theta_{k}}{2})e^{-{\it i}\varepsilon_{k}\tau(t)}+\sin^{2}(\frac{\theta_{k}}{2})e^{{\it i}\varepsilon_{k}\tau(t)},
 \\
u_{k,12}(t)
&=-u_{k,21}(t)
 \\
&=
{\color{black}-}{\it i}\cos(\frac{\theta_{k}}{2})\sin(\frac{\theta_{k}}{2})(e^{{-\it i}\varepsilon_{k}\tau(t)}-e^{{\it i}\varepsilon_{k}\tau(t)}),
 \\
u_{k,22}(t)
&=
\sin^{2}(\frac{\theta_{k}}{2})e^{-{\it i}\varepsilon_{k}\tau(t)}+\cos^{2}(\frac{\theta_{k}}{2})e^{{\it i}\varepsilon_{k}\tau(t)}.
\quad\quad
 \\
\el
\end{eqnarray}
%%%%%%%%%%%%%%%%%%%%%%%%%%%%%%%%%%%%%%%%%%%%%%%%%%%%%%%%%%%%%%%%%%%%%%%
Here $\theta_k$ is the polar angle that parametrizes $\boldsymbol{n}_k$. Combining Eqs. (\ref{eq:rhok})-(\ref{eq:Uii}), a straightforward but lengthy calculation yields a closed formula
 for the mixed-state Loschmidt amplitude of the $k$:th mode,
%%%%%%%%%%%%%%%%%%%%%%%%%%%%%%%%%%
\begin{eqnarray}
 \label{eq:exact}
 \bl
{\cal G}^{\rho}_k(t) =
\cos(\varepsilon_{k}\tau(t))
{\color{black} +} {\color{black} i\sin(\varepsilon_{k}\tau(t))\tanh(\lambda(0)\beta\varepsilon_{k})}.
\quad
\el
\end{eqnarray}
%%%%%%%%%%%%%%%%%%%%%%%%%%%%%%%%%%
When ${\cal G}^{\rho}_k(t) \!= \!0$, the mixed state rate function $g^{\rho}(t) \!= \!-(1/N)\ln {\cal L}^{\rho}(t) $ of the Loschmidt echo (with ${\cal L}^{\rho}(t) = \prod_k |{\cal G}^{\rho}_k(t)|^2$) becomes nonanalytic, signaling a DQPT.
As seen from Eq. (\ref{eq:exact}), ${\cal G}^{\rho}_k(t)$ has zeros only in the infinite-temperature limit $\beta \rightarrow 0$, {\color{black} for which it reduces to the pure-state single-mode Loschmidt amplitude in Eq. (\ref{eq:cosine})}. This is expected from our treatment of pure states: Only  critical modes, occupying the upper and lower bands with equal probability, can make a Loschmidt amplitude vanish for certain values of $t$ (the critical times). When $\beta \rightarrow 0$ {\em all} modes of the system become critical since in this limit the thermal ensemble is maximally mixed. One concludes, and checks with Eq. (\ref{eq:exact}), that the model now exhibits a vast number of sequences of DQPTs, happening at critical times $t_{k\!,n}^{\ast}$ satisfying
%%%%%%%%%%%%%%%%%%%%%%%%%%%%%%%%%%
\begin{equation} \label{eq:kcritical}
\tau(t_{k\!,n}^{\ast}) = \frac{\pi}{\varepsilon_{k}}(n+1/2);\;
 n= 0,1,2,\ldots
\end{equation}
%%%%%%%%%%%%%%%%%%%%%%%%%%%%%%%%%%
with $k$ taking {\em all} allowed values in the Brillouin zone.
\\

Admittedly, initializing a state by coupling a system to a heat bath in the infinite-temperature limit $\beta \rightarrow 0$ is not a very realistic option for an experimentalist! However, the reasoning can be sharpened by tweaking it slightly.

First, let us assume that the system is isolated from its environment on experimentally relevant time scale (using, e.g., an ultracold atomic gas trapped in an optical lattice). Next, let us also assume that one may be able to realize a fully connected {\em nonintegrable} mode coupling, making all modes coupled to all others. This kind of setup, in configuration space, has been conjectured to create fast scrambling~\cite{Lashkari2013,Bentsen2019} $-$ a process where local information spreads over a quantum system on a time scale which is only logarithmic with the size of the system. Fast scrambling is expected to asymptotically lead to maximal mixing of the modes, with the reduced density matrix $\rho_k$ of each mode evolving into a thermal state with an {\em effective} infinite temperature. (Note that the assumption of nonintegrability is here crucial, since otherwise a generalized Gibbs ensemble may ensue~\cite{Vidmar2016}.)  Having reached maximal mixing, within experimental limits, the mode coupling may then be turned off, while simultaneously, at the reference time $t=0$, the synchronized periodic drive is turned on, now predicted to cause DQPTs under time evolution.

Fast scrambling may be only one among several {\color{black} possible} routes for realizing DQPTs  under synchronized periodic driving {\color{black} (or, more precisely, precursors of DQPTs under synchronized periodic driving, since any experimental system is finite)}. In Sec. V we shall discuss another possible protocol for initializing a maximally mixed state of the system $-$ maybe easier to realize in the laboratory $-$ using slow periodic driving~\cite{Dalessio2014,Lazarides2014,Bukov2016}, again with the degrees of freedom coupled by a nonintegrable interaction, but now not required to be fully connected.

Our discussion $-$ whether about pure or mixed states $-$ has been patterned on the simple example of two-band fermion models. As we have already mentioned, these are most often equated with tight-binding lattice models in an independent-particle approximation or with mean-field superconductors. However, none of these types of systems are very likely to be realizable under synchronized periodic driving. More viable candidates are spin systems, where two-band fermionic models enter the stage {\color{black} as effective} Hamiltonians governing the spin dynamics. We shall study one such model in the next section, in part as a preparation for the discussion in Sec. IV, but also to unveil several more intriguing features of synchronized periodic driving.
\\ \\ \\

\section{Case study: anisotropic XY chain \\ in a transverse magnetic field \label{SE3}}

We have chosen as benchmark model the anisotropic XY chain in a transverse magnetic field under synchronized periodic driving. The static model is defined by the Hamiltonian
%%%%%%%%%%%%%%%%%%%%%%%%
\begin{equation}\label{eq:XY}
H \!=\!-\frac{J}{2}\sum\limits_{n = 1}^N {\left( {(1\! + \!\gamma )\sigma _n^x \sigma_{n \!+ \!1}^x \!+\! (1 \!- \!\gamma ) \sigma_n^y \sigma_{n \!+\! 1}^y} \right)} \!-\! h\sum\limits_{n \!= \!1}^N {\sigma_n^z},
\end{equation}
%%%%%%%%%%%%%%%%%%%%%%
with $N$ the number of sites on the chain, and where $\sigma_{\alpha} with \alpha=x,y,z$ are the Pauli matrices. $J>0$ denotes the ferromagnetic exchange coupling, and $h$ and $\gamma$ are the magnitude of the magnetic field and the anisotropy parameter, respectively. We here consider the case with periodic boundary conditions, $\sigma^{\alpha}_{n+N} = \sigma^{\alpha}_n$.

The {\color{black} XY chain} defined by Eq. (\ref{eq:XY}) is one of the best studied lattice spin models. It exhibits an equilibrium quantum phase transition in the thermodynamic limit $N \rightarrow \infty$ at a critical magnetic field $h_c = J$, from a ferromagnetic phase ($h<J$) to a paramagnetic phase ($h>J$). The case $\gamma=1$ yields the transverse field Ising chain, arguably the simplest {\color{black} instance} exhibiting such a transition~\cite{Sachdev}.

The exact solution of the XY chain is well known~\cite{Lieb1961} and we here only recap the basic steps. The Hamiltonian in Eq. (\ref{eq:XY}) is first mapped onto a second-quantized Hamiltonian ${\cal H}$ of free spinless fermions by a Jordan-Wigner transformation,
%%%%%%%%%%%%%%%%%%%%%%%%%%%%
%%%%%%%%%%%%%%%%%%%%%%%%%%%%%%%%%%%%%%%%%%%%%%%%%%%%
\bea
\no
\sigma^{+}_{n}&=& \sigma^{x}_{n} + {\it i}\sigma^{y}x_{n}=\prod_{m=1}^{n-1}(1-2c_{m}^{\dagger}c_{m})c_{n}^{\dagger},\\
\sigma^{-}_{n}&=& \sigma^{x}_{n} - {\it i}\sigma^{y}_{n} = \prod_{m=1}^{n-1}c_{n}(1-2c_{m}^{\dagger}c_{m}),\\
\no
\sigma^{z}_{n} &=& c_{n}^{\dagger}c_{n} -\frac{1}{2},
\eea
%%%%%%%%%%%%%%%%%%%%%%%%%%%%%%%%%%%%%%%%%%%%%%%%%%%
with $c_n^{\dagger}$ ($c_n$) fermion creation (annihilation) operators. We have here dropped a boundary term induced
by the Jordan-Wigner transformation since this term contributes only to
${\cal O}(1/N)$ in the energy spectrum and hence is negligible for large $N$. Fourier transforming the operators,
$c_n = \sum_k e^{-ikn}c_k$, and introducing a Nambu spinor ${\boldsymbol{c}}^{\dagger}_{k}\!=\!(c_{k}^{\dagger},~c_{-k})$,  ${\cal H}$ can be decomposed as ${\cal H} =\sum_{k} {\boldsymbol{c}}^{\dagger}_{k}H_{k}{\boldsymbol{c}}_{k}$, where (choosing to work in the even parity sector of the fermionic Fock space~\cite{Franchini2017}) the sum runs over half the Brillouin zone with $k = \pi/N, 3\pi/N,\ldots,(N-1)\pi/N$. The first-quantized mode Hamiltonian $H_k$ is given by
%%%%%%%%%%%%%%%%%%%%%%%%%%%%%%%%%%%%%%%%%  Eq. 40  %%%%%%%%%%%%%%%%%%%%%%%%%%%%%%%%%%%%%%%%%%%
\bea
\label{eq:bandXY}
H_{k}=2\left(
\begin{array}{cc}
-h_{z}(k) & -{\it i}h_{xy}(k) \\
{\it i}h_{xy}(k) & h_{z}(k)\\
\end{array}
\right)
\eea
%%%%%%%%%%%%%%%%%%%%%%%%%%%%%%%%%%%%%%%%%%%%%%%%%%%%%%%%%%%%%%%%%%%%%%%%%%%%%%%%%%%%
with $h_z(k) = J\cos(k) + h$ and $h_{xy}(k)= J\gamma\sin(k)$. The spectrum is now easily obtained by a Bogoliubov transformation, and one finds
%%%%%%%%%%%%%%%%%%%%%%%%%%%%%%%%%%%%%%%  Eq. 41  %%%%%%%%%%%%%%%%%%%%%%%%%%%%%%%%%%%%%%%%%%%%%
\begin{equation} \label{eq:XYenergy}
\varepsilon_{k}^{\pm}=\pm 2\sqrt{P^{2}(k)+Q^{2}(k)}=\pm\varepsilon_{k},
\end{equation}
%%%%%%%%%%%%%%%%%%%%%%%%%%%%%%%%%%%%%%%%%%%%%%%%%%%%%%%%%%%%%%%%%%%%%%%%%%%%%%%%%%%%%%%%
with eigenstates
%%%%%%%%%%%%%%%%%%%%%%%%%%%%%%%%%%%%  Eq. 6   %%%%%%%%%%%%%%%%%%%%%%%%%%%%%%%%%%%%%%%%%%%%%%
\bea
\label{eq:eigenstates}
|\varphi^{+}_{k}\rangle\!=\!\left(
                          \begin{array}{c}
                             {\it i}\sin(\frac{\theta_{k}}{2}) \\
                             \cos(\frac{\theta_{k}}{2}) \\
                          \end{array}
                        \right), \ \
|\varphi^{-}_{k}\rangle\!=\!\left(
                          \begin{array}{c}
                            \cos(\frac{\theta_{k}}{2}) \\
                             -{\it i}\sin(\frac{\theta_{k}}{2}) \\
                          \end{array}
                        \right),
                        \quad
\eea
%%%%%%%%%%%%%%%%%%%%%%%%%%%%%%%%%%%%%%%%%%%%%%%%%%%%%%%%%%%%%%%%%%%%%%%%%%%%%%%%%%
where $P(k)=J\cos(k)+h$, $Q(k)=\gamma J\sin(k)$, and {\color{black} $\theta_{k}=-\arctan(Q(k)/P(k))$}. The ground state $|\psi_0\rangle$ of the system is obtained as in Eq. (\ref{eq:product}): by filling all single-particle states $|\varphi^{-}_{k}\rangle$ of the lower band. To investigate the topology of the ground state it is convenient to got to a Majorana representation (``Kitaev chain"). For details, and a discussion also of (nonsynchronized) periodically driven Kitaev chains, we refer the reader to Ref. \cite{Chen2018}.

\subsection{Pure state dynamical quantum phase transitions\label{SE3a}}

As follows from the discussion in Sec. II, the time evolution of the eigenstates in (\ref{eq:eigenstates}) under a synchronized drive $\lambda(t)$ is simply given by $|\varphi^{\pm}_{k}(t)\rangle=e^{\mp{\it i}\varepsilon_{k}\tau(t)}|\varphi^{\pm}_{k}\rangle$ where $\tau(t)=\int_{0}^{t}\lambda(t')dt'$. Thus, if the system is prepared in the ground state at $t=0$, the Loschmidt echo ${\cal L}(t)$ {\color{black} takes the value unity} and there is no DQPT. Only if the initial state contains a critical mode $k^{\prime}$ occupying the lower and upper band with the same probability will the drive set off a sequence of DQPTs at critical times $t_{k^{\prime}\!,n}^{\ast}$, {\color{black} cf. Eq. (\ref{eq:tau}).}

{\color{black} In close analogy to an equilibrium phase transition, classical or quantum, a DQPT appears only in the thermodynamic limit since the rate function $g_{\cal G}(t)$ of the Loschmidt amplitude in Eq. (\ref{eq:rate}) is analytic for any finite $N$~\cite{Heyl2018,Zvyagin2016}. The exponential suppression of the Loschmidt amplitude for large $N$, implied by Eq. (\ref{eq:rate}), therefore makes a direct observation of a DQPT difficult. At a practical level, a precursor of a DQPT in a large but finite system, signaled by the emergence of cusplike features in $g_{\cal G}(t)$ (or in $g(t) = 2$Re$[g_{\cal G}(t)]$, the associated rate function of the Loschmidt echo), may still be challenging to pick up in an experiment or a computer simulation: A DQPT present due to a {\em single} critical mode $k'$ at a time $t_{k^{\prime}}^{\ast}$ becomes visible only if the experiment or the computation can resolve  $t_{k^{\prime}}^{\ast}$ to a precision such that $-\ln{\cal G}(t_{k^{\prime}}^{\ast}) \gg N$. Else the cusplike feature gets washed out by the averaging over $N$ in the formula for the rate function, Eq. (\ref{eq:rate}). The problem is evaded if there is a number of critical modes $\{k^{\prime}\}, \,\sim N$, with the same (or nearly the same) critical time(s), together surviving the averaging over $N$. In the present context of the effective fermionic two-band model that emulates the XY chain, this requires two conditions to be fulfilled. First, the system must be initialized with a sufficient number of modes occupying the lower and upper band with the same probability. Secondly, to share the same critical time(s), the energies
$\varepsilon_{k^{\prime}}$ in Eq. (\ref{eq:XYenergy}) can depend only weakly on $k^{\prime}$ (requiring that $J/h \ll 1$), since otherwise the critical times for the different modes will disperse; cf Eq. (\ref{eq:kcritical}). In the following we explore the scenario when both these conditions are satisfied, with special attention to effects from synchronized periodic driving.}

\subsubsection{Critical times}

%%%%%%%%%%%%%%%%%%%%%%%  Fig.1   %%%%%%%%%%%%%%%%%%%%%%%
\begin{figure*}[t]
\begin{minipage}{\linewidth}
\centerline{\includegraphics[width=\linewidth]{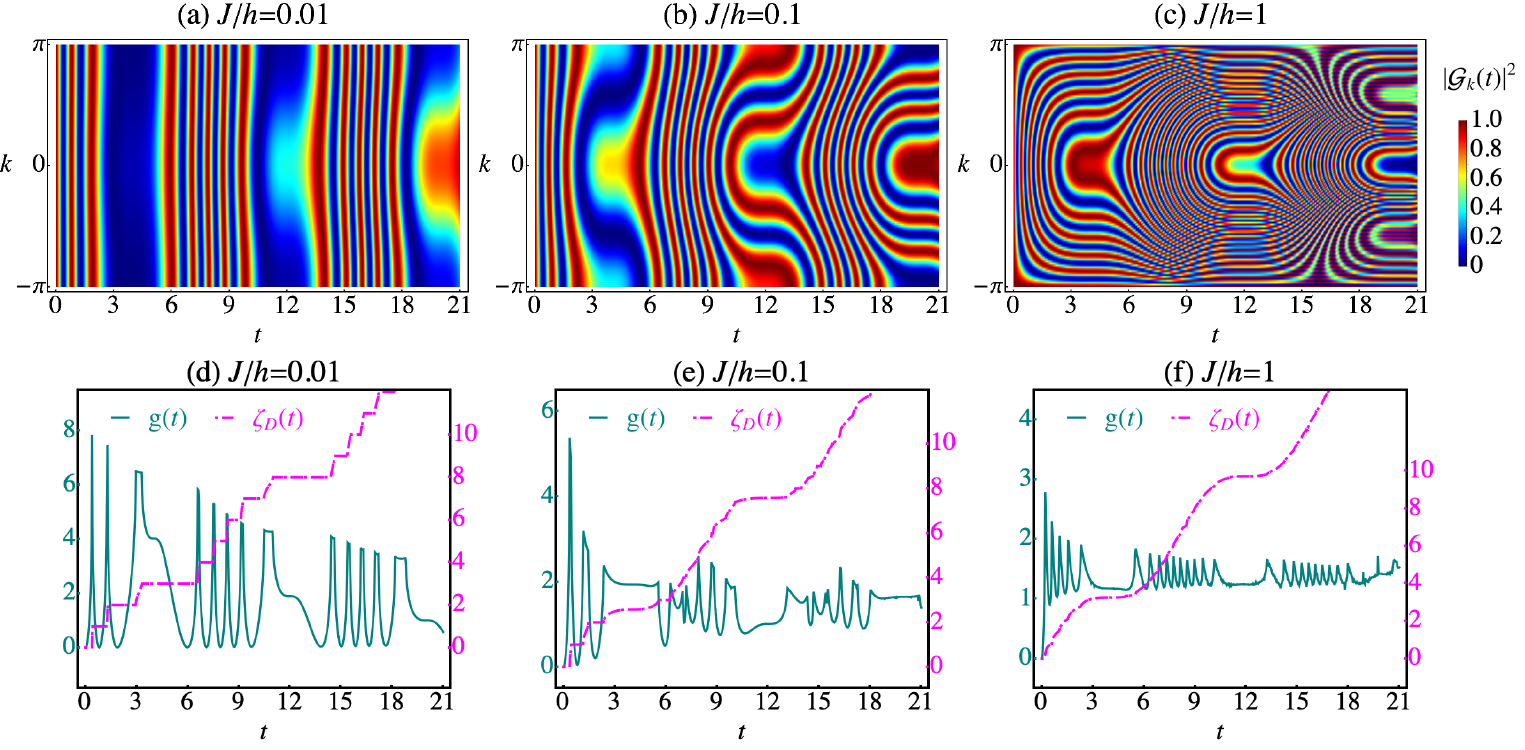}}
\centering
\end{minipage}
%===============================================================
\caption{(Color online)
{\color{black} The upper row of panels show contour plots of the single-mode Loschmidt echo $|{\cal G}_k(t)|^2$ of the anisotropic XY chain in a transverse field when $\gamma = 1$ {\em (transverse field Ising chain)} as a function of momentum $k$ and time $t$ for (a) $J/h=0.01$; (b) $J/h=0.1$; and (c)
$J/h=1$, with synchronized periodic drive $\lambda(t) = 1 + \cos(\pi t/4)$. The lower row of panels, (d)-(f), display the corresponding time dependence of the Loschmidt echo rate function $g(t)$ and the effective dynamical order parameter $\zeta_D(t)$.}}
\label{fig1}
\end{figure*}
%%%%%%%%%%%%%%%%%%%%%%%%%%%%%%%%%%%%%%%%%%%%%%%%%%%%%%%
For simplicity, we shall study the case with the system initialized with {\em all} modes $n = 1, 2,\ldots, N$ occupying the lower and upper band with the same probability, predicted to yield an abundance of DQPTs appearing at critical times $t_{k\!,n}^{\ast}$ with $k = \pm \pi/N, \pm 3\pi/N,\ldots, \pm (N-1)\pi/N$. For pure states this is a Gedanken experiment only, useful to highlight certain features of DQPTs under synchronized driving. Moreover, it provides an instructive background to a discussion of such DQPTs with mixed states, a scenario which may actually be possible to realize in the laboratory.

With this protocol, the Loschmidt amplitude is given by
%%%%%%%%%%%%%%%%%%%%%%%%%%%%%%%%%%%%%%%
\begin{equation} \label{eq:Gprotocol}
{\cal G}(t) = \prod_k {\cal G}_k(t) = \prod_{k} \langle \varphi_k | \varphi_k(t)\rangle,
\end{equation}
%%%%%%%%%%%%%%%%%%%%%%%%%%%%%%%%%%%%%%%
where
%%%%%%%%%%%%%%%%%%%%%%%%%%%%%%%%%%%%%%
\begin{equation} \label{eq:initial}
|\varphi_k(t)\rangle = \frac{1}{\sqrt{2}}\left(\mbox{e}^{i\varepsilon_k\tau(t)}|\varphi_k^{-}\rangle + \mbox{e}^{-i\varepsilon_k\tau(t)}|\varphi_k^{+}\rangle\right)
\end{equation}
%%%%%%%%%%%%%%%%%%%%%%%%%%%%%%%%%%%%%%%
with $\varepsilon_k$ and $|\varphi_k^{\pm}\rangle$ to be read off from Eq. (\ref{eq:XYenergy}) and (\ref{eq:eigenstates}), respectively.

In Fig.~\ref{fig1}(a)-(c) we display the Loschmidt echo ${\cal L}_k(t)$ in color-coded contour plots for a chain with $N \rightarrow \infty$ sites, choosing $\gamma = 1$ (a.k.a. the  transverse field Ising chain) with three different values of $J/h$. The driving function is taken to be the same as in our discussion in Sec. \ref{SE2a}:
$\lambda(t) = \lambda_0 + \lambda_1\cos(\omega t)$, implying that $\tau(t) = \int_0^t \lambda(t^\prime) dt^\prime = \lambda_0 t + (\lambda_1/\omega) \sin(\omega t)$, {\color{black} where in Fig.~\ref{fig1} we have chosen $\lambda_0\!=\!\lambda_1\!=\!1$ and $\omega\!=\!\pi/4$.}

Let us inspect panel (a). The zeros $t_{k\!,n}^{\ast}$ of ${\cal G}_k(t)$, and hence those of the Loschmidt echo ${\cal L}_k(t) = |{\cal G}_k(t)|^2$, are here seen to coalesce at a few critical times, giving rise to the near-vertical black streaks in the contour plot. The associated nonanalyticities in the rate function $g(t) = 2\mbox{Re}[g_{\cal G}(t)]$ of the Loschmidt echo show up as distinct cusps in Fig.~\ref{fig1}(d), with
%%%%%%%%%%%%%%%%%%
\begin{eqnarray} \label{eq:rate2}
g(t) &=& -\lim_{N\rightarrow \infty} \frac{1}{N} \sum_{k} \ln {\cal L}_k(t) \nonumber \\
& = & - \frac{1}{2\pi} \int_{-\pi}^{\pi} \ln {\cal L}_k(t) dk,
\end{eqnarray}
%%%%%%%%%%%%%%%%%%
as follows from Eqs. (\ref{eq:rate}) and (\ref{eq:Gprotocol}). Notably, {\color{black}Fig.~\ref{fig1}(d) reveals that the sequence of the individual cusps, and hence the DQPTs, are manifestly {\color{black} nonperiodic. Note} that the amplitudes of the peaks in $g(t)$ are seen to decay with time, as caused by the linear term $\lambda_0t$ in $\tau(t)$ which disperses the critical momenta in Eq. (\ref{eq:kcritical}) as $t$ increases.}

Panels (b) and (c) of Fig.~\ref{fig1} show contour plots where the zeros of ${\cal L}_k(t)$ form increasingly snake-like structures as $J/h$ gets larger. The diminished clustering at a few critical times as compared to (a) implies that the peaks of the rate function $g(t)$ in {\color{black} panels (e) and (f)} now have lower heights. Note also that there is no signal in Fig.~\ref{fig1} that the equilibrium quantum critical point at $J/h =1$ plays any particular role; what matters and what controls the qualitatively different behaviors is only the magnitude of $J/h$. The feature that the DQPT critical times become only very weakly $k$-dependent when $J/h$ is tuned to a small value (as in panel (a)) is easy to see by inserting the expression for $\varepsilon_k$ in (\ref{eq:XYenergy}) into Eq. (\ref{eq:tau}). One may speculate that this reflects a tendency towards a certain rigidity of the system when the magnetic field $\sim h$ dominates the spin exchange $\sim J$, enforcing a strong spin polarization in the ground state of the (static) XY chain.

\begin{figure*}[t]
\begin{minipage}{\linewidth}
\centerline{\includegraphics[width=\linewidth]{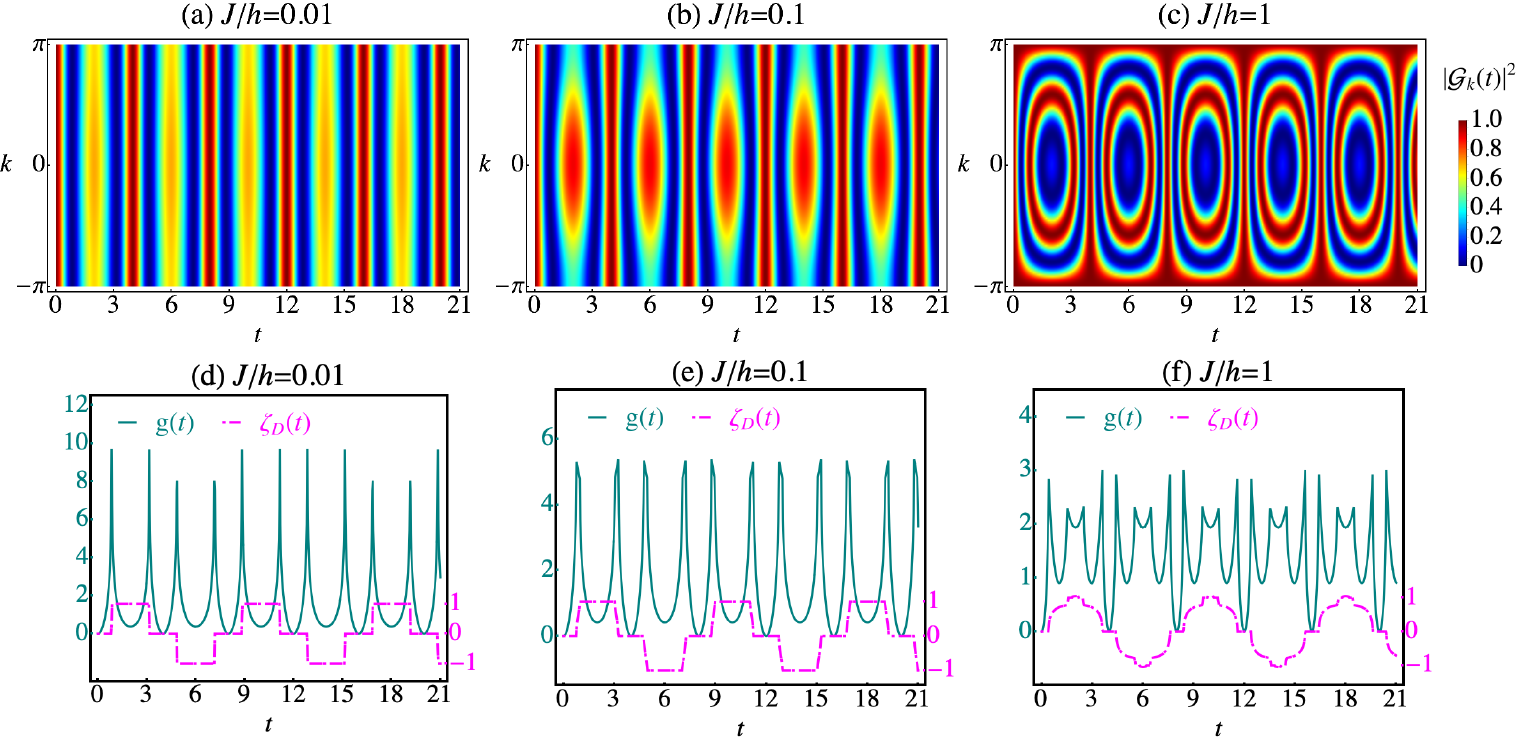}}
\centering
\end{minipage}
%===============================================================
\caption{(Color online)
{\color{black} Plots of the Loschmidt echo $|{\cal G}_k(t)|^2$, rate function $g(t)$, and effective dynamical order parameter $\zeta_D(t)$ for the transverse field Ising chain with the same parameter values as in Fig.~1, but with the offset in the synchronized periodic drive removed, i.e., with $\lambda(t) = \cos(\pi t/4)$.}}
\label{fig2}
\end{figure*}

{\color{black} The contour plots depicted in Fig.~\ref{fig1} will change with a change of the driving function. Keeping to the same harmonic modulation as above but removing the offset $\lambda_0$, i.e., with    $\lambda(t) = \lambda_1\cos(\omega t)$, yields the contour plots in Fig.~\ref{fig2}(a)-(c). Having removed the offset, the periodicity of $\tau(t) = (1/\omega)\sin(\omega t)$ is now seen to bring about a periodicity in the overall pattern of the Loschmidt echo. This, in turn, gives rise to clusters of DQPTs (signaled by the cusps in panels (d)-(f) of Fig.~\ref{fig2}) which repeat periodically with no attenuation of the rate function $g(t)$ with time. One notes in particular that these features are insensitive to the choice of the ratio $J/h$, with $J/h$ only setting the scale of the rate function.}

%%%%%%%%%%%%%%%%%%%%%%%  Fig.2   %%%%%%%%%%%%%%%%%%%%%%%
%%%%%%%%%%%%%%%%%%%%%%%%%%%%%%%%%%%%%%%%%%%%%%%%%%%%%%%

\subsubsection{Effective dynamical order parameter}

As was first shown in Ref.~\cite{Budich2016}, dynamical phases separated by DQPTs triggered by a single quantum quench may be characterized by a dynamical topological order parameter. It was later found that this holds also for periodically driven systems, whether the DQPTs are set off by a quench that repeats periodically~\cite{Zhou2021} or by a protocol which allows the system to be described by a time-independent Hamiltonian in a rotating frame~\cite{yang2019floquet}.

This raises the question whether the same is true also for dynamical phases in systems under synchronized periodic driving. The answer is in the negative: Since the existence of DQPTs in these systems relies on a fine tuning of the initial state, there can be no topological invariant that provides protection of its dynamical phases. Still, it is instructive to understand, at a formal level, {\em how} a dynamical topological order parameter fails to capture the physics under a synchronized drive. This will lead us to suggest a definition of an alternative {\em effective} order parameter under synchronized periodic driving, useful on short and intermediate time scales given certain conditions. As we shall find for our benchmark model $-$ the anisotropic XY chain in a transverse magnetic field $-$ the important condition is that the ground state of the {\color{black} initial} (static) Hamiltonian is strongly spin polarized.

The construction of a dynamical topological order parameter takes off from the identification of a quantity that behaves discontinuously if and only if there is a DQPT, i.e., at a zero of the Loschmidt amplitude. In Ref.~\cite{Budich2016}, the Pancharatnam geometric phase $\phi^G(t)$~\cite{Pancharatnam1956} $-$ which generalizes the Berry phase to noncyclic (and not necessarily adiabatic) time evolutions of a state vector~\cite{Samuel1988} $-$ was put forward as a suitable choice. $\phi^G(t)$ is defined as the gauge invariant component of the phase $\phi(t)$ of the Loschmidt amplitude, ${\cal G}(t) = |{\cal G}(t)|\mbox{e}^{i\phi(t)}$, obtained by subtracting the dynamical phase
$\phi^{\text{dyn}}(t)=-i\int_0^t \langle \psi(t')| \partial_{t^{\prime}} |\psi(t')\rangle dt'$ from $\phi(t)$:
%%%%%%%%%%%%%%%%%
\begin{equation} \label{eq:PGP}
\phi^G(t) = \phi(t) - \phi^{\text{dyn}}(t).
\end{equation}
%%%%%%%%%%%%%%%%
Considering a particle-hole symmetric two-band fermionic mode Hamiltonian $H_k$, the Pancharatnam phase $\phi^G(k,t)$ of a state vector labelled by $k$ vanishes for $k=0,\pi$, implying that the map $k \in [0,\pi] \rightarrow \mbox{e}^{i{\phi}^G(k,t)}$ can be characterized by an integer-valued winding number
%%%%%%%%%%%%%%%
\begin{equation} \label{eq:winding}
\nu_{{D}}(t) = \frac{1}{2\pi}\oint_0^{\pi} \frac{\partial \phi^G(k,t)}{\partial k} dk.
\end{equation}
%%%%%%%%%%%%%%%
This topological invariant defines a dynamical topological order parameter, which, in analogy to order parameters of equilibrium topological phase transitions, jumps by an integer at a DQPT and specifies the dynamical phase in between two such transitions.

Let us now turn to our synchronously driven two-band model that represents the XY chain, with single-mode Loschmidt amplitudes ${\cal G}(k,t) = |{\cal G}(k,t)|\mbox{e}^{i\phi(k,t)} = \langle \varphi_k | \varphi_k(t)\rangle$. Here $| \varphi_k(t)\rangle$ is given in Eq. (\ref{eq:initial}). {\color{black} As follows from Eq. (\ref{eq:cosine}),
%%%%%%%%%%%%%%%%%%%%%%%%%%%%%%%% Eq.30 %%%%%%%%%%%%%%%%%%%%%%%%%%%%%%%%
\begin{eqnarray}
\label{eq:totalphase}
\phi(k,t)=\mbox{arg}\big(\!\cos(\varepsilon_k\tau(t))\big),
\end{eqnarray}
%%%%%%%%%%%%%%%%%%%%%%%%%%%%%%%%%%%%%%%%%%%%%%%%%%%%%%%%%%%%%%%%%%%%
with $\varepsilon_k$ given in Eq. (\ref{eq:XYenergy}). The phase $\phi(k,t)$ is seen to be quantized with $\pi$-phase slips at the critical times $t_{k,n}^{\ast}$ in Eq. (\ref{eq:tau}), formally a consequence of the staircase structure of the arg function with a cosine as argument, implying that $\phi(k,t) = n\pi$ for ${\color{black}(n-1/2)\pi < \varepsilon_k \tau(t) < (n+1/2)\pi,}\ n=0,\pm 1, \pm 2,\ldots$.} Considering the dynamical phase $\phi^{\text{dyn}}(k,t)$ for a single mode, it can be written as
%%%%%%%%%%%%%%%%%%%%%%%%%%%%%%%%%%%%%%%%%%%%%%%%
\begin{equation} \label{eq:dynphase}
\phi^{\text{dyn}}(k,t) = -i \int_0^t \langle \varphi_k(t^{\prime}) | \frac{d \tau}{dt^{\prime}} \partial_{\tau} | \varphi_k(t^{\prime}) \rangle dt^{\prime},
\end{equation}
%%%%%%%%%%%%%%%%%%%%%%%%%%%%%%%%%%%%%%%%%%%%%%%
recalling that the state $|\varphi_k(t^{\prime})\rangle$ depends on time via the integral $\tau(t^{\prime})$ of the driving function; cf. text after Eq. (\ref{eq:1}).
Inserting the expression for $|\varphi_k(t^{\prime})\rangle$ using Eq. (\ref{eq:initial}) , one verifies that $\phi^{\text{dyn}}(k,t)$ in (\ref{eq:dynphase}) vanishes identically, independent of the choice of the driving function, and by that, independent of the specifics of $\tau(t^{\prime})$. It follows from Eq. (\ref{eq:PGP}) that
%%%%%%%%%%%%%%%%
\begin{equation} \label{eq:equality}
\phi^G(k,t) = \phi(k,t),
\end{equation}
%%%%%%%%%%%%%%
implying that $\phi^G(k,t)/\pi$ is quantized, taking integer values which jump by unity at a DQPT. In particular, this means that the dynamical topological order parameter $\nu_D(t)$ in Eq. (\ref{eq:winding}) vanishes identically.

The $\pi$-phase slips of the single-mode Pancharatnam phase $\phi^G(k,t)$ at the critical times $t^{\ast}_{k,n}$ suggests that it may still serve as a basis for a differently constructed dynamical order parameter, devoid of topology, but still useful for characterizing a dynamical phase between two DQPTs. Naming it $\zeta_D(t)$, we define it as the average of $\phi^G(k,t)/\pi$,
%%%%%%%%%%%%%%%
\begin{equation} \label{eq:DOP}
\zeta_D(t) = \frac{1}{2\pi^2} \int_{-\pi}^{\pi} \phi^G(k,t) dk,
\end{equation}
%%%%%%%%%%%%%%
assuming, as above, that $N$ is sufficiently large to allow for an integration over the Brillouin zone.

As an illustration, let us again consider the driving function $\lambda(t) = \lambda_0 + \lambda_1\cos(\omega t)$, with $\tau(t) = \lambda_0 t + (\lambda_1/\omega) \sin(\omega t)$. By inspection of the expression for $\varepsilon_k$ in Eq. (\ref{eq:XYenergy}), one notes that $\varepsilon_k \tau(t)$ will be only weakly $k$-dependent for small $J/h$ unless $t$ takes large values. In other words, within these constraints $-$ a small ratio $J/h$ and not too large time series $-$ $\phi^G(k,t)/\pi$ will take the {\em same} integer value for {\em all} momenta $k$ except when getting very close to a branch cut (a.k.a. a critical time) where the small spread of $\varepsilon_k \tau(t)$ for different $k$ will make $\phi^G(k,t)/\pi$ jump to the next integer at slightly different $k$-dependent critical times {\color{black} (cf. text after Eq. (\ref{eq:totalphase}))}. It follows that $\zeta_D(t)$ in Eq. (\ref{eq:DOP}) will exhibit a staircase structure, as seen in panel (d) of Fig.~\ref{fig1}, but with the vertical steps slightly tilted due to the narrowly shifted jumps for different modes. {\color{black} The same behavior of                $\zeta_D(t)$ is manifest in panel (d) of Fig.~\ref{fig2}, now with driving function $\lambda(t) = \lambda_1\cos(\omega t)$. Note, however, that while the vertical steps remain well defined up to small tilts (like in Fig.~\ref{fig1}(d)), the staircase in Fig.~\ref{fig2}(d) is not ascending since the linear term $\lambda_0 t$ is absent from $\tau(t)$.}
%Another difference, not visible on the time scale displayed in Figs.~\ref{fig1} and \ref{fig2}, is that the staircase in \ref{fig2}(d) does not get corrupted as time passes:
%The function $\tau(t)$ is periodic and  preserves the weak $k$-dependence of the Pancharatnam phase with time (for the small value of $J/h$ in this figure).}

{\color{black} The quantity $\zeta_D(t)$ for small values of $J/h$, as in Figs.~\ref{fig1}(d) and \ref{fig2}(d),} serves much the same purpose as the dynamical topological order parameter $\nu_{\text{D}}(t)$:  $\zeta_D(t)$ signals a DQPT (but now with a limited resolution) and assigns an integer number to the dynamical phase in between two such transitions. Since this number is not a winding number, however, but {\color{black} simply an average over $\phi^G(k,t)/\pi$, it should be defined mod $2$ (since $\phi^G(k,t)$ is an angle {\color{black} and given by the principal value of the arg function in Eq. (\ref{eq:totalphase})}), suggesting that the dynamical phases come in only two classes, labelled by $\zeta_D(t) \!=\! 0$ and $1$. One may be tempted to speculate that this feature may point to the {\color{black} presence} of an emergent $Z_2$ symmetry of the driven XY chain. However, such a conclusion would be premature. The quantity $\zeta_D(t)$ can only stand in for an {\em effective} order parameter since it exhibits a {\color{black} continuous change} between 0 and 1 when going from one phase to the next $-$ albeit extremely fast for the cases in Figs.~\ref{fig1}(d) and \ref{fig2}(d), producing slightly tilted vertical steps in the staircases. Since a {\color{black} ramp, however steep}, does not signal a change of a discrete symmetry, such a change can be argued for only by constructing an improved dynamical order parameter, if this is at all possible.

{\color{black} As $J/h$ and (or) time $t$ get(s) larger, the ramps in $\zeta_D(t)$ become more pronounced, with the vertical steps in a staircase getting more tilted and the horizontal steps more narrow, rendering $\zeta_D(t)$ impractical as an effective order parameter.} Eventually the staircase structure is all but washed out and replaced by a {\color{black} smooth graph (after coarse graining over short time intervals)}; cf. panels (e) and (f) in Figs.~\ref{fig1} and \ref{fig2}. Fig.~\ref{fig3} vividly illustrates why the {\color{black} Pancharatnam phase $\phi^G(k,t)$ for different values of $J/h$ yield such different behaviors of $\zeta(t)$ when the driving function has an offset, like for the case displayed in Fig.~\ref{fig1}.}

%%%%%%%%%%%%%%%%%%%%%%%  Fig.3  %%%%%%%%%%%%%%%%%%%%%%%
\begin{figure}[t]
\begin{minipage}{\columnwidth}
\centerline{\includegraphics[width=\columnwidth]{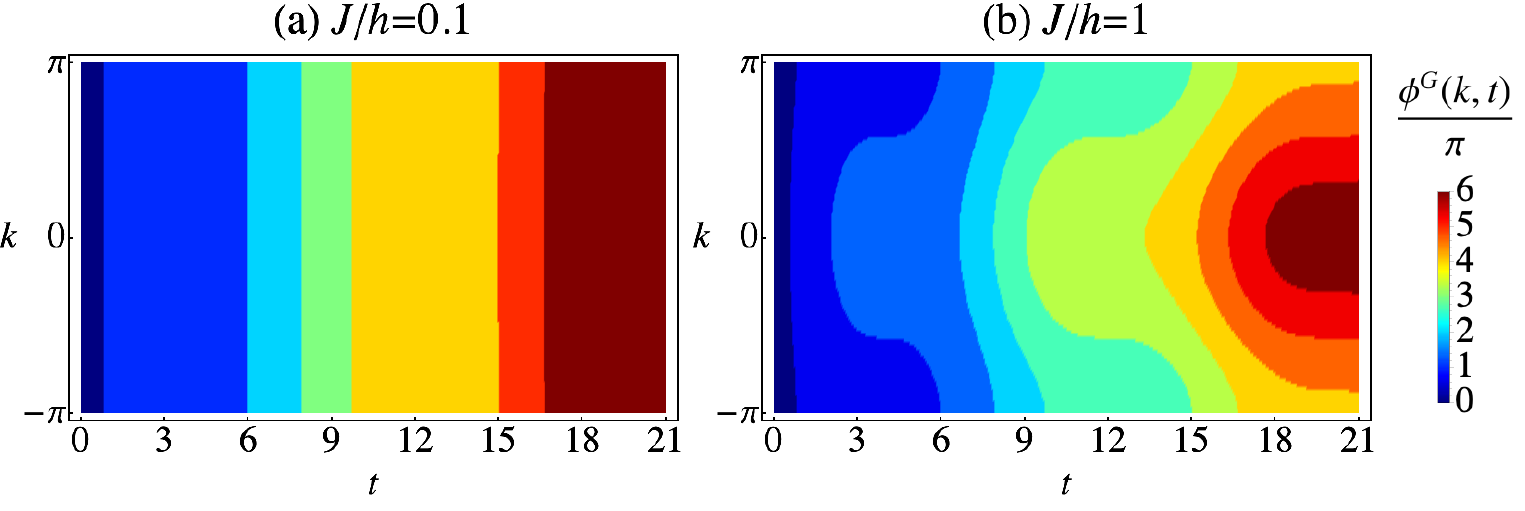}}
\centering
\end{minipage}
%===============================================================
\caption{(Color online) {\color{black} Contour plots of the single-mode geometric Pancharatnam phase $\phi^{G}(k,t)$ as function of momentum $k$ and time $t$ for the transverse field Ising chain with synchronized periodic drive $\lambda(t) = 1 + \cos(\pi t/4)$ when (a) $J/h=0.1$; (b) $J/h=1$. }}
\label{fig3}
\end{figure}
%%%%%%%%%%%%%%%%%%%%%%%%%%%%%%%%%%%%%%%%%%%%%%%%%%%%%%%

One may inquire how the tuning of other parameters than $J/h$ and $\lambda_0/\lambda_1$ influences the appearance of DQPTs in the synchronously driven XY chain. Numerical data for a large range of choices of the XY anisotropy $\gamma$ and the frequency $\omega$ of the drive show that changing the values of these parameters do not make a qualitative change compared to the cases that we have discussed in this section \cite{Numerical}. What matters for the overall picture of DQPTs are the ratios $J/h$ and $\lambda_0/\lambda_1$, where the cusps in the rate function become increasingly sharper and repeat for longer times the smaller one (or both) of these ratios is (are) taken $-$ with {\color{black} the clusters of peaks in the rate function $g(t)$} being periodic and persistent in time only when $\lambda_0/\lambda_1 = 0$.

More pressing is the question what happens when deviating slightly from the {\color{black} fine-tuned} $t\!=\!0$ initial state in Eq. (\ref{eq:initial}). While the pristine DQPTs in the thermodynamic limit $-$ manifested by visible nonanalyticities in the dynamical free energy of each single $k$-mode $-$ are killed off as soon as the amplitudes for the two eigenstates $|\varphi_k^{\pm}\rangle$ in Eq. (\ref{eq:initial}) become different, this is not obvious for the signatures of the DQPTs captured in Figs.~\ref{fig1} and \ref{fig2}, emerging from a collective clustering of single-mode nonanalyticities around shared critical times. Maybe these signatures will still be present, albeit somewhat blurred, {\color{black} serving as precursors of DQPTs}? Rather than addressing this question head on, we defer it to the next section in the context of mixed states $-$ with mixed states being more relevant for discussing possible experimental realizations.

\subsection{Mixed state dynamical quantum phase transitions\label{SE3b}}

%%%%%%%%%%%%%%%%%%%%%%%  Fig.4  %%%%%%%%%%%%%%%%%%%%%%%
\begin{figure*}[t]
\begin{minipage}{\linewidth}
\centerline{\includegraphics[width=\linewidth]{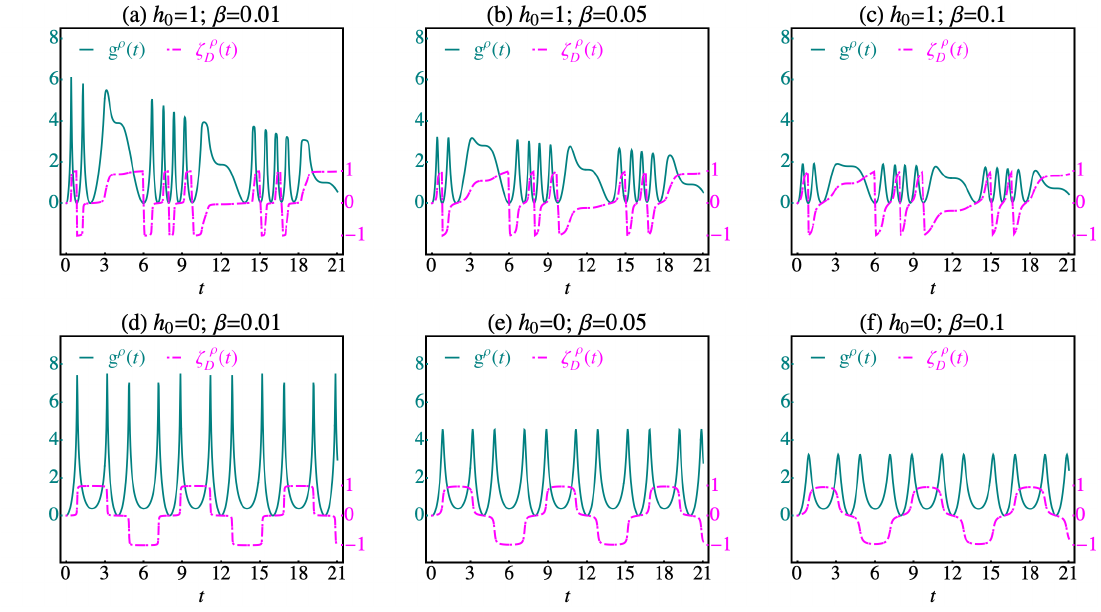}}
\centering
\end{minipage}
%===============================================================
\caption{(Color online)
The upper row of panels show plots of the time dependence of the mixed-state Loschmidt echo rate function $g^{\rho}(t)$ and the mixed-state effective dynamical order parameter $\zeta^{\rho}_{D}(t)$ for the transverse field Ising chain with {\color{black} $J/h=0.01$ and} synchronized periodic drive $\lambda(t)= 1 + \cos(\pi t/4)$ at inverse temperature (a) $\beta = 0.01$; (b) $\beta = 0.05$; and (c)  $\beta = 0.1$. The lower row of panels, (d)-(f), show the corresponding time dependence when the offset in the drive is removed, i.e., with $\lambda(t)=\cos(\pi t/4)$.}
\label{fig4}
\end{figure*}
%%%%%%%%%%%%%%%%%%%%%%%%%%%%%%%%%%%%%%%%%%%%%%%%%%%%%%%

In Sec. (\ref{SE2b}) we laid out a scheme for analyzing DQPTs in fermionic two-band systems under {\color{black} synchronized} periodic driving, with the systems initialized in a mixed state.
Specifically, we obtained an exact formula for the single-mode mixed-state Loschmidt amplitude ${\cal G}^{\rho}_k(t)$, Eq. (\ref{eq:exact}), in terms of $\lambda(0), \tau(t),\varepsilon_k$, and the polar angle $\theta_k$ (with the latter defined in the text after Eq. (\ref{eq:Hk0})). Thus, knowing $\lambda(0)$ and $\tau(t)$, an application of Eq. (\ref{eq:exact}) to the anisotropic XY chain in a transverse field only requires to collect the appropriate expressions for $\varepsilon_k$ and $\theta_k$, with $\varepsilon_k$ given by Eq. (\ref{eq:XYenergy}), and where $\theta_k$ is obtained by inserting $H_k$ from Eq. (\ref{eq:bandXY}) into Eq. (\ref{eq:Hk0}).

\subsubsection{Critical times}
{\color{black} Given the expression for ${\cal G}^{\rho}_k(t)$ in Eq. (\ref{eq:exact}), with parameters adapted to the XY chain,} we numerically plot the rate function $g^{\rho}(t)$ of the mixed-state Loschmidt echo ${\cal L}_{\rho}(t) = \prod_k {\cal L}^{\rho}_k(t) = \prod_k |{\cal G}_k^{\rho}|^2$, {\color{black} defined as in Eq. (\ref{eq:rate2}) with $g^{\rho}(t)$ and ${\cal L}^{\rho}_k(t)$ replacing $g(t)$ and ${\cal L}_k(t)$, respectively.}  Taking the same synchronized drives and parameter values as in the preceding section, and choosing $\beta =0$ (i.e., {\color{black} effective} infinite temperature), one obtains plots identical with those in Figs.~\ref{fig1} and \ref{fig2} {\color{black} as realized by simply comparing Eq. (\ref{eq:exact}) with (\ref{eq:totalphase}).}  As we have discussed, this is anticipated since in this case the reduced density matrices for the states of the decoupled $k$-modes are maximally mixed: The Loschmidt echo {\color{black} does not discriminate between maximal mixing and equal-amplitude pure-state superpositions of two orthogonal states.} Being an expected outcome at infinite temperature, we refrain from plotting this particular case.

To address the question whether there are still signatures of DQPTs present when deviating slightly from a maximally mixed initial state, we have plotted the rate function $g^{\rho}(t)$ of the Loschmidt echo for $\beta = 0.01, 0.05,$ and $0.1$ in Fig.~\ref{fig4}(a)-(c). {\color{black} We here consider the case when the driving function $\lambda(t)$ has an offset, like in Fig.~1. Further, }we have chosen the same ratio $J/h=0.01$ as in Fig.~\ref{fig1}(d) (favoring strong spin polarization in the ground state of the static model), the case in Fig.~1 with the most distinct cusps in the rate function. As evidenced by panel (a) of Fig.~\ref{fig4}, the signatures of DQPTs are still manifest {\color{black} when $\beta$ is taken sufficiently small}, but then taper off as $\beta$ increases (panel (b) and (c)).

{\color{black} Removing the offset from the driving function yields the rate function plots in Fig.~\ref{fig4}(d)-(f). The overall picture is much the same as with a nonzero offset: strong signatures of DQPTs in panel (d) with the same critical times as in the pure state case of Fig.~\ref{fig2}(d), with a softening of the cusps as $\beta$ increases (panel (e) and (f)).}

\subsubsection{Effective dynamical order parameter}

{\color{black} The finite-temperature counterpart of the effective dynamical order parameter $\zeta_D(t)$ discussed in the previous section can be defined in analogy to
Eq. (\ref{eq:DOP}), as an average of $\phi_{\rho}^G(k,t)/\pi$,
%%%%%%%%%%%%%%%%%%%%%%%%%%%%
\begin{equation} \label{feq:finiteTorder}
\zeta^{\rho}_D(t) = \frac{1}{2\pi^2}\int_{-\pi}^{\pi} \phi_{\rho}^G(k,t) dk,
\end{equation}
%%%%%%%%%%%%%%%%%%%%%%%%%%%
with $\phi_{\rho}^G(k,t)$ the mixed-state Pancharatnam phase.
As shown by Sj\"oqvist {\em et al.}~\cite{Sjoqvist2000}, the expression for $\phi_{\rho}^G(k,t)$,
%%%%%%%%%%%%%%%%%%%%%%%%%%%%%
\begin{equation} \label{eq:finiteTgeometric}
\phi_{\rho}^G(k,t) = \phi_{\rho}(k,t) - \phi_{\rho}^{\text{dyn}}(k,t),
\end{equation}
%%%%%%%%%%%%%%%%%%%%%%%%%%%%%
is obtained by extending the construction of the Pancharatnam phase for pure states, Eq. (\ref{eq:PGP}), to purifications of mixed states, Eq. (\ref{eq:pure}).
The phase $\phi_{\rho}(k,t)$ is that of the mixed-state Loschmidt amplitude ${\cal G}_k^{\rho}(t) = |{\cal G}_k^{\rho}(t)|\exp(i\phi_{\rho}(k,t))$ and can be read off directly from Eq. (\ref{eq:exact}):
%%%%%%%%%%%%%%%%%%%%%%%%%%%%%
%%%%%%%%%%%%%%%%%%%%%%%%%%%%%
%\begin{equation} \label{eq:XYfiniteTLophase}
%\phi_{\rho}(k,t) \!=\! \arctan
%\Big[
%\frac{P^2(k)\!-\!Q^2(k)}{P^2(k)\!+\!Q^2(k)}
%\tan[
%\varepsilon_k\tau(t)
%]
%\tanh[
%\lambda(0)\beta\varepsilon_k
%]
%\Big],
%\end{equation}
%%%%%%%%%%%%%%%%%%%%%%%%%%%%%
%%%%%%%%%%%%%%%%%%%%%%%%%%%%%
{\color{black}
\begin{equation} \label{eq:XYfiniteTLophase}
\phi_{\rho}(k,t) =
\arg\Big(\!\cos\big(\varepsilon_k \tau(t)\big)
  \!+\! i
  \sin\big(\varepsilon_k \tau(t)\big)
  \tanh\big(
  \lambda(0)\beta\varepsilon_k\big)\Big).
\end{equation}
}
%%%%%%%%%%%%%%%%%%%%%%%%%%%%%
The mixed-state dynamical phase $\phi_{\rho}^{\text{dyn}}(k,t)$ in turn is given by
%%%%%%%%%%%%%%%%%%%%%%%%%%%%%
\begin{equation} \label{eq:finiteTdynphase}
\phi_{\rho}^{\text{dyn}}(k,t)=-\int_{0}^{t} \mbox{Tr}\Big(\rho_{k}(t')H_{k}(t')\Big)dt',
\end{equation}
%%%%%%%%%%%%%%%%%%%%%%%%%%%%%
using the definition $\phi^{\text{dyn}}(t)=-i\int_0^t \langle w(t')| \partial_{t^{\prime}} | w(t')\rangle dt'$ applied to the purified state $|w(t)\rangle = U_k(t,0)|w(0) \rangle$, cf. Eq. (\ref{eq:pure}) (where, in Eq. (\ref{eq:finiteTdynphase}), $\rho_k(t)$ is the time-evolved density matrix). An explicit expression for $\phi_{\rho}^{\text{dyn}}(k,t)$ for the XY chain is obtained from Eq. (\ref{eq:finiteTdynphase}) using Eqs. (\ref{eq:rhok})-(\ref{eq:exact}):
%%%%%%%%%%%%%%%%%%%%%%%%%%%%%
\begin{equation} \label{eq:fXYfiniteTdynphase}
\phi_{\rho}^{\text{dyn}}(k,t)={\color{black} -}\tanh\big(\lambda(0)\beta\varepsilon_k)\big)
{\color{black} \varepsilon_k} \tau(t).
\end{equation}
%%%%%%%%%%%%%%%%%%%%%%%%%%%%%

{\color{black} In Fig \ref{fig4}, superposed on the rate function graphs, we have plotted $\zeta_D^{\rho}(t)$ for three effective temperatures close to maximal mixing. Taken together,} the results displayed in Fig.~\ref{fig4} suggest that signatures of precursors of DQPTs under synchronized periodic driving may actually be picked up in an experiment $-$ with these signatures due to collective clustering of a large number of critical modes. {\color{black} While amplitude and sharpness of the signatures degrade as one departs from maximal mixing, their very presence suggest that a DQPT can be inferred from data away from (effective) infinite temperature.} This should put less demand on the preparation of the initial state of the system. In the next section we elaborate on this and related issues with a view to possible experimental observations. {\color{black} Before concluding this section, let us mention that an alternative approach to interpreting zeros of Loschmidt amplitudes for mixed states has been proposed in Ref.~\cite{Hou2020}, taking off from the Uhlmann geometric phase~\cite{Uhlmann1986}.}

\section{Towards experimental observations}

Experimental breakthroughs with analog quantum simulators have allowed observations of quench-induced DQPTs {\color{black} exploiting a variety of platforms: trapped ions~\cite{Jurcevic2017,Zhang2017} and Rydberg atoms~\cite{Bernien2017}, cold atomic gases~\cite{Flaschner2018,YangTianYang2019,Tian2020}, superconducting qubits~\cite{Guo2019}, photonic quantum walks~\cite{wang2019simulating}, nanomechanical oscillators~\cite{Tian2019}, nuclear spins~\cite{Nie2020}, and NV centers~\cite{Chen2020}.} What are the prospects for experimental observations of the type of DQPTs that we have studied here? Without going into details, let us outline some of the opportunities and challenges.

The required experimental backbones are threefold: (i) realization/simulation of a synchronously driven system; (ii) preparation of a nearly maximally mixed initial state; and (iii) detection of DQPT signatures.
\\

(i) {\em Realization/simulation} $-$ Taking off from the realm of the anisotropic XY chain, there have been a number of experiments probing the dynamics of short transverse field Ising chains simulated by trapped ions~\cite{Friedenauer2008,Lanyon2011} however, with finite-range interactions between the spins. While this feature may not impact the appearance of DQPTs, one would ideally opt for a platform that can simulate nearest-neighbor spin interactions in larger chains, and crucially, which is also sufficiently versatile to allow for synchronized periodic driving of the spin coupling and the magnitude of the transverse magnetic field. A setup that comes across as particularly promising for this purpose is the analog spin chain simulator recently proposed by Nguyen {\em et al.}~\cite{Nguyen2018}, based on laser-trapped circular Rydberg atoms. The proposed device realizes a spin-1/2 XXZ Hamiltonian in a magnetic field, with fully tunable parameters, allowing for a realization of the transverse field Ising chain with nearest-neighbor interactions. Importantly, all parameters can easily be modulated in time within a wide range of frequencies, in principle allowing for synchronized driving. Thus, as for the problem of realization/simulation, prospects look bright.
\\

(ii) {\em Initial state preparation} $-$ In Sec. \ref{SE2b} we briefly discussed a scheme to prepare a (near) maximally mixed initial state by exploiting fast scrambling of spins~\cite{Lashkari2013,Bentsen2019}. However, experimental studies of spin scrambling have only recently come of age~\cite{Garttner2017}, making this route somewhat uncertain. More critically, fast scrambling appears incompatible with the circular Rydberg simulator in Ref.~\cite{Nguyen2018} as it assumes spin interactions of uniform strength ranging over the full system. A simpler scenario, more likely to be realizable within the setting of this device, is to add a single magnetic impurity to the XXZ chain. As shown recently by Brenes {\em et al.}~\cite{Brenes2020}, such a local perturbation of the integrable XXZ Hamiltonian gives rise to eigenstate thermalization. By this, the system is expected to quickly heat up close to an infinite-temperature state when subject to a slow periodic drive~\cite{Dalessio2014,Lazarides2014,Bukov2016}. Provided that this kind of perturbation can be mimicked by inserting a defect into the chain of Rydberg atoms, this would be a way to experimentally prepare a(n almost) maximally mixed state. Having obtained this state, the tunable parameters are then set to the transverse field Ising Hamiltonian and the synchronized drive is turned on.

The viability of this type of scenario is for the experimentalist to judge. Quite likely, there may be other, maybe more direct ways to tweak the circular Rydberg simulator so as to achieve nonintegrability, and by that, near maximal mixing of the initial state.
\\

(iii) {\em Detection} $-$ Direct detection of DQPTs requires experimental access to the Loschmidt echo or some related quantity, like the quantum work statistics~\cite{Silva2008}, from which the rate function $g(t)$ can be deduced. An experimental protocol for measuring the Loschmidt echo for interacting Rydberg atoms in a lattice has been proposed in Ref.~\cite{Macri2016}, suggesting that it is in principle accessible with the circular Rydberg simulator proposed in Ref. \cite{Nguyen2018}. Other observables, indirectly linked to signatures of DQPTs, have been proposed \cite{Nicola2021,Bandyopadhyay2021,Halimeh2021} or put to actual use in recent experiments \cite{Jurcevic2017}. A specific proposal for detecting Floquet DQPTs under synchronized periodic drives has recently been put forward by one of us \cite{Zamani2022}, based on the notion of out-of-time-order correlations. 
Measurements of such correlations have been reported in the literature \cite{Li2017}, however in the different context of a nuclear magnetic resonance quantum simulator, and not with periodic drives. The prospect that similar measurements could be carried out also for the Rydberg simulator with an applied synchronized drive is attractive but needs to be demonstrated in the laboratory. For a theoretical discussion we refer the reader to Ref. \cite{Zamani2022}. 
\\

All things considered, the three legs (i)-(iii) of a required experiment are expected to be demanding to carry out, however, there is a robust platform to build on $-$ the circular Rydberg simulator put forward by Nguyen {\em et al.}~\cite{Nguyen2018}. This gives good grounds to anticipate that DQPTs under synchronized periodic driving will {\color{black} soon become} accessible in the laboratory.

\section{Summary}

{\color{black} To summarize, we have analyzed the occurrence of Floquet DQPTs in a quantum system subject to synchronized periodic driving, more specifically, DQPTs in a generic 1D fermionic two-band model with all terms in its Hamiltonian subject to harmonic periodic drives with the same {\color{black} frequency and phase}. As known from general theory~\cite{Heyl2018,Huang2016}, all DQPTs are conditioned on the presence of ``critical modes" which occupy the available energy levels with equal probability, i.e., in the language of an open system, modes in maximally mixed states. This can happen dynamically, in topologically protected DQPTs, or, in ``accidental" DQPTs by fine tuning the Hamiltonian or the initial condition. Taking the effective model of the anisotropic XY chain in a transverse magnetic field as a benchmark $-$ with Jordan-Wigner fermions emulating the spin dynamics $-$ we have examined the case where the system is initialized with {\em all} modes in a maximally mixed state (or, very close to maximal mixing), causing a coalescence of the critical times for the modes, in effect producing  ``collective"  DQPTs (or, precursors of such DQPTs) provided that the magnetic field is large compared to the energy scale of the spin exchange.

Our analysis shows that the existence of a DQPT can be inferred from numerical data for the rate function of the Loschmidt echo also when the initial state of the system is perturbed slightly away from maximal mixing. This is somewhat analogous to the well known situation where a continuous equilibrium phase transition can be inferred from finite-size data (with a system off criticality), with precise predictions about scaling, universality, and critical exponents obtainable via finite-size scaling~\cite{Fisher1972}. It would be interesting to find 
a well-controlled procedure which would allow precise characteristics of DQPTs to be predicted from off-critical data (coming, e.g., from the blurring of an initial condition as in Sec. III.B), 
similar in spirit to how finite-size scaling allows properties of equilibrium phase transitions to be deduced from off-critical data.

A synchronized periodic drive is found to give rise to nonperiodic sequences of {\color{black} DQPTs with an attenuation} of the amplitude of the rate function of the Loschmidt echo with time when there is an initial offset of the harmonic drive (causing a sudden change of the scale of the Hamiltonian {\color{black} when the drive is turned on}). If there is no offset, clusters of DQPTs {\color{black} repeat periodically} with no attenuation, allowing for the monitoring of DQPTs in long time series. The nonperiodic pattern of DQPTs {\em within} a cluster can be calculated exactly, knowing the driving frequency and the band structure of the model. This offers {\color{black} an expedient route to a thorough comparison between theory and} experiment. With the current rapid advances in realizing analog quantum simulators, such a test may not be too far into the future. } \\ \\

\section*{Acknowledgements} This work was supported by the Swedish Research Council under grant no. 621-2014- 5972.
A.~A. acknowledges the support of the Max Planck- POSTECH-Hsinchu Center for Complex Phase
Materials, and financial support from the National Research Foundation (NRF) funded by the Ministry
of Science of Korea (Grant No. 2016K1A4A01922028).

\bibliography{Refs}

\newpage

\end{document}